\documentclass[a4j,11pt,onecolumn,oneside,notitlepage,final]{article}
\title{}
\author{}
\date{\today}
\usepackage{amsmath}
\usepackage{amssymb}
\usepackage{amsfonts}

\usepackage{graphicx}
\usepackage{bm}
\usepackage[doublespacing]{setspace}

\usepackage{natbib}
\usepackage{comment}
\usepackage{color}
\usepackage[top=30truemm,bottom=30truemm,left=25truemm,right=25truemm]{geometry}

\usepackage{arydshln}
\usepackage{natbib}

\newcommand{\C}{\mathbb{C}}

\begin{document}

\title{
Pair circulas modelling for multivariate circular time series\thanks{The research reported herein was supported by JSPS KAKENHI Grant Numbers 	18K11193.}
}
\author{Hiroaki Ogata}

\maketitle

\begin{abstract}
Modelling multivariate circular time series is considered. The cross-sectional and serial dependence is described by circulas, which are analogs of copulas for circular distributions. In order to obtain a simple expression of the dependence structure, we decompose a multivariate circula density to a product of several pair circula densities. Moreover, to reduce the number of pair circula densities, we consider strictly stationary multi-order Markov processes. The real data analysis, in which the proposed model is fitted to multivariate time series wind direction data is also given.  
%Some simulation studies are provided to see the behavior of the proposed model.
\end{abstract}

\section{Introduction}\label{seq:Introduction}

Circular data stands for the data that take its value on a unit circle. The typical example is a directional data. The direction is expressed as an angle from a certain origin point, and it can be represented by a point on a unit circle. 
The directional data is observed in many research field. The wind direction is observed in environmental science or meteorology, the movement direction of a certain animal is recorded in biology, the direction of river flow is considered in geography, to name a few. 
Due to its periodic feature, we cannot apply the usual statistical techniques, such as arithmetic mean, multiplication etc., to the circular data. Therefore, the specific treatment ought to be applied. As introductions of the circular statistics, we can refer to \cite{jammalamadaka2001topics} and \cite{MJ09}, for example.   

In practice, the circular data is often the time series data, too. The wind direction data observed during a certain time period is one of the examples. For statistical modelling of circular time series data, \cite{breckling1989analysis} introduced the von Mises process and the wrapped autoregressive process.  \cite{fisher1994time} proposed the projected Gaussian process and the processes derived using link functions.  
\cite{wehrly1980bivariate} proposed the stationary circular Markov process and \cite{abe2017circular} elucidated the structure of the circular autocorrelation function of this model. \cite{ogata2023mixture} extended the circular Markov model considered in \cite{abe2017circular} to a multi-order circular Markov process, in which the conditional distribution given by all past values depends on not only the previous value but also several adjacent past values.  

More generally, we can consider the multivariate circular time series data, such as the wind direction data at several different points observed during a certain time period.   
In this paper, we attempt to model the  multivariate circular time series. It is accomplished by use of the circula, which is an analog of the copula for circular distribution. The concept of this circular version copula originally come from \cite{wehrly1980bivariate}, and it was named circula by \cite{jones2015class}. In order to enable the simple interpretation of the dependence structure, we decompose the multivariate circula to a product of several pair circulas. This technique is similar to that of vine copulas.
Readers can refer, e.g., \cite{Joe1996}, \cite{BC01} and \cite{ACFB09} on several types of vine copulas and related works. 
In addition, we assume the strictly stationary multi-order circular Markov process to make the model not too large. For linear random variables, \cite{BS15} and \cite{S15} consider the vine copula specifications for stationary multivariate Markov chains and stationary multivariate Markov processes, respectively. 
  
% In fact, the circular Markov process of \cite{wehrly1980bivariate} is constructed by two arbitrary circular density functions, and one of them corresponds to a copula density function in linear random variables. 
%\cite{jones2015class} called this circular version copula density function 'circula'. The construction of multi-order circular Markov process is accomplished by using a pair-circula decomposition of a multivariate circular distribution. 

The paper is organized as follows. The construction of multi-order circular Markov process by use of pair circulas is described in Section \ref{sec:Model}. In Section \ref{sec:Real data analysis}, 
%several second-order circular Markov processes are generated and their circular autocorrelations are calculated in order to demonstrate the properties of the processes. 
we fit the second-order circular Markov process to real wind direction data observed at different three locations, and estimate parameters by MCMC. 
%Section \ref{sec:Conclusion and future research} concludes the findings of this work and provides future research topics. 

\section{Model}\label{sec:Model}
Let $ \bm{\Theta}_t=(\Theta_{1,t}, \ldots, \Theta_{m,t})^\prime \in \C^m $ be a continuous-valued circular random vector, observed at times $ t=1, \ldots, T $, where $ \C $ denote a unit circle. Put the all elements in one column and denote it as $ \bm{\Theta}=(\bm{\Theta}_1^\prime, \ldots, \bm{\Theta}_T^\prime)^\prime \in \C^{Tm}$.  
For simplicity, we also use another expression $ \bm{\Phi} = (\Phi_1, \ldots, \Phi_N)^\prime = \bm{\Theta} $ with $ N=Tm $, that is, $ \Theta_{j,t} = \Phi_{(t-1)m+j} $.

For modelling multivariate circular time series, we use the circulas, which are analogs of copulas for circular distributions. A general $ N $-variate circular density $f$ on $ \C^N $ can be written by its circula density $ c $ on $ \C^N $ and its marginal circular density and distribution functions $ f_j $, $ F_j $ $ (j=1, \ldots, N) $ on $ \C $, as
\begin{align}\label{eq:circular joint density}
f(\bm{\phi})=(2\pi)^N c( \upsilon_1, \ldots, \upsilon_N ) \prod_{i=1}^{N} f_i(\phi_i). 
\end{align}
Here, $ \bm{\upsilon} 
%= (\bm{\upsilon}^\prime_1, \ldots, \bm{\upsilon}^\prime _T) 
= (\upsilon_1, \ldots, \upsilon_N)^\prime = (2\pi F_1(\phi_1), \ldots, 2\pi F_N(\phi_N))^\prime = (\bm{\psi}_1^\prime, \ldots, \bm{\psi}_T^\prime)^\prime = \bm{\psi}$ where $\bm{\psi}_t = (\psi_{1,t}, \ldots, \psi_{m,t})^\prime$ is a vector for time $t$. The all marginal distributions of circula are circular uniform distributions, that is 
\begin{align*}
c(\upsilon_j) = \int_{[0,2\pi)^{N-1}} c(\upsilon_1, \ldots, \upsilon_N) \  d\bm{\upsilon}_{-j} = \frac{1}{2\pi}
\end{align*}
where 
%$ \Pi $ indicates the interval $ [0,2\pi) $, and 
$ \bm{\upsilon}_{-j} $ indicates the vector excluding $ j $th component from $ \bm{\upsilon} $. Furthermore, the circula density requires the periodicity:
\begin{align*}
c(\upsilon_1+2k_1\pi, \ldots, \upsilon_N+2k_N\pi) = c(\upsilon_1, \ldots, \upsilon_N), \quad k_j \in \mathbb{Z} \ \mbox{for} \ j=1, \ldots, N.
\end{align*}
Due to this periodicity, the circula is different from the rescaled linear copula density. 
The bivariate case is introduced in \cite{jones2015class}. The expression (\ref{eq:circular joint density}) is its direct extension to general $ N $-dimensional case. 

Hereafter, for $ a \ge b $, $ \bm{\upsilon}_{a:b} $ indicates the vector $ (\upsilon_a, \upsilon_{a-1}, \ldots, \upsilon_b)^\prime $ and $ \bm{\upsilon}_{-(a:b)} $ indicates the vector, excluding $ \bm{\upsilon}_{a:b} $ from $ \bm{\upsilon} $. The joint circula density of $ \bm{\upsilon}_{a:b} $ is
\begin{align*}
c(\upsilon_{a:b}) = \int_{[0,2\pi)^{N-a+b-1}} c(\bm{\upsilon}) \ d\bm{\upsilon}_{-(a:b)}
\end{align*} 
is also a circula density. Following the manner of \cite{S15}, '$ c $' and '$ C $' exclusively denote density functions and distribution functions of circulas. 

For $ a-1 \ge b+1 $, the conditional circula density of $ \upsilon_a $ given $ \bm{\upsilon}_{a-1:b+1} $, the conditional circula density of $ \upsilon_b $ given $ \bm{\upsilon}_{a-1:b+1} $, and conditional joint circula density of $ \upsilon_a $ and $ \upsilon_b $ given $ \bm{\upsilon}_{a-1:b+1} $ are
\begin{align}
&f(\upsilon_a | \bm{\upsilon}_{a-1:b+1})=\frac{c(\bm{\upsilon}_{a:b+1})}{c(\bm{\upsilon}_{a-1:b+1})}, \label{eq:a}\\ 
&f(\upsilon_b | \bm{\upsilon}_{a-1:b+1})=\frac{c(\bm{\upsilon}_{a-1:b})}{c(\bm{\upsilon}_{a-1:b+1})}, \label{eq:b}\\
&f(\upsilon_a, \upsilon_b | \bm{\upsilon}_{a-1:b+1})=\frac{c(\bm{\upsilon}_{a:b})}{c(\bm{\upsilon}_{a-1:b+1})} \label{eq:ab}
\end{align} 
respectively. Note that these conditional circula densities are not circula densities any more because their marginals are not circular uniform. Corresponding conditional circula distribution functions of (\ref{eq:a})-(\ref{eq:ab}) are denoted by $ F(\upsilon_a | \bm{\upsilon}_{a-1:b+1}) $, $ F(\upsilon_b | \bm{\upsilon}_{a-1:b+1}) $ and $ F(\upsilon_a, \upsilon_b | \bm{\upsilon}_{a-1:b+1}) $, respectively. When $ a-1 < b+1 $, we treat them as unconditional ones. For example, $ f(\upsilon_a | \bm{\upsilon}_{a-1:b+1}) = f(\upsilon_a) = c(\upsilon_a) = (2\pi)^{-1}$. Strictly speaking, these functions themselves should be also indexed, like $ c_{a:b}(\bm{\upsilon}_{a:b}) $, $ f_{a|a-1:b+1}(\upsilon_a | \bm{\upsilon}_{a-1:b+1}) $, $ f_{a,b|a-1:b+1}(\upsilon_a, \upsilon_b | \bm{\upsilon}_{a-1:b+1}) $, and so on, but we omit them for simple expressions. 

Now, we consider to decompose the $ N $-dimensional circula density function $ c $ into several pair circula density functions $ c^\ast_{i,j} $ as 
\begin{align}\label{eq:pair circulas decomposition}
c(\bm{\upsilon})
=(2\pi)^{-N} \prod_{i=2}^{N} \prod_{j=1}^{i-1} (2\pi)^2 c^\ast_{i,j}\bigl(2\pi F(\upsilon_i|\bm{\upsilon}_{i-1:j+1}), 2\pi F(\upsilon_j|\bm{\upsilon}_{i-1:j+1})\bigl).
\end{align}
The calculation for (\ref{eq:pair circulas decomposition}) is given in Appendix. By this decomposition, we can easily interprete the dependence structure among the variables.  

These $ N(N-1)/2 $ pair circulas can be grouped together into some blocks in order to distinguish the serial and cross-sectional dependence. For different time points $ t_1 > t_2 $, define
\begin{align}\label{eq:serial dependence}
K_{t_1,t_2}(\bm{\psi}_{t_2}, \ldots, \bm{\psi}_{t_1})
=\prod_{i=a(t_1)}^{b(t_1)} \prod_{j=a(t_2)}^{b(t_2)} 
(2\pi)^2 c^\ast_{i,j}\bigl(2\pi F(\upsilon_i|\bm{\upsilon}_{i-1:j+1}), 2\pi F(\upsilon_j|\bm{\upsilon}_{i-1:j+1})\bigl) 
\end{align}
and for same time point $ t $, define
\begin{align}\label{eq:cross-sectional dependence}
K_{t,t}(\bm{\psi}_{t})
=\prod_{i=a(t)+1}^{b(t)} \prod_{j=a(t)}^{i-1} 
(2\pi)^2 c^\ast_{i,j}\bigl(2\pi F(\upsilon_i|\bm{\upsilon}_{i-1:j+1}), 2\pi F(\upsilon_j|\bm{\upsilon}_{i-1:j+1})\bigl). 
\end{align}
Here, $ a(t)=(t-1)m+1 $ and $ b(t)=tm $. (\ref{eq:serial dependence}) and (\ref{eq:cross-sectional dependence}) describe serial and cross-sectional dependence, respectively. Then, the joint circula $ c(\bm{\upsilon}) $ can be expressed as 
\begin{align}\label{eq:block copula}
c(\bm{\upsilon})
=c(\bm{\psi}_1, \ldots, \bm{\psi}_T)
=\left\{ \prod_{t=1}^{T} K_{t,t}(\bm{\psi}_{t}) \right\}
\left\{ \prod_{t=2}^{T} \prod_{i=1}^{t-1} K_{t,t-i}(\bm{\psi}_{t-i}, \ldots, \bm{\psi}_{t}) \right\}.
\end{align}

When $ T $ is large, the model is too huge to handle because it includes $ N(N-1)/2 $ different pair circulas. The remedy for it is to assume the strictly stationary structure. If the process $ \{\bm{\Theta}_t \}_{t =1, \ldots, T} $ is strictly stationary, the depenence structure between $ \bm{\Theta}_t $ and $ \bm{\Theta}_{t-k} $ depends on only the time difference $ k $. Therefore, the expressions $ K_{t,t} $ and $ K_{t,t-k} $ in (\ref{eq:block copula}) become $ K_{0} $ and $ K_{k} $, respectively. In addition, if we assume the $ p $-th order Markov structure, $ \bm{\Theta}_t $ and $ \bm{\Theta}_{t-k} $ becomes independent when $ k > p $, which leads to $ K_{k} = 1 $ for $ k > p $. In sum, if $ \{\bm{\Theta}_t \}_{t =1, \ldots, T} $ is strictly stationary $ p $-th order Markov process, the expression of joint circula $ c(\bm{\upsilon}) $ in (\ref{eq:block copula}) becomes
\begin{align}\label{eq:pair circulas decomposition in stationary Markov}
c(\bm{\upsilon})
=c(\bm{\psi}_1, \ldots, \bm{\psi}_T)
=\left\{ \prod_{t=1}^{T} K_{0}(\bm{\psi}_{t}) \right\}
\left\{ \prod_{t=2}^{T} \prod_{k=1}^{\min(t-1,p)} K_{k}(\bm{\psi}_{t-k}, \ldots, \bm{\psi}_{t}) \right\},
\end{align}
which contains $ m(m-1)/2+m^2p $ possibly different pair circula density functions.

\subsection{Constructing pair circulas}
\citet{jones2015class} proposes a way to construct the pair circulas. Let $ \Theta_1 $ follow the circular uniform distribution and $ \Omega $ follow a circular distribution with density $ g $, independently of $ \Theta_1 $. If we define $ \Theta_2=\Omega+q\Theta_1 $, where $ q \in \{ -1, 1 \} $ is non-random, then $ \Theta_2 $ follows the circular uniform distribution and the conditional density of $ \Theta_2 | \Theta_1=\theta_1 $ is $ g(\theta_2-q\theta_1) $. This means we can construct a  circula density which links $\Theta_1 $ and $ \Theta_2 $ as
\begin{align}\label{eq:construction a pair circula}
c(\theta_1, \theta_2) = \frac{1}{2 \pi} g(\theta_2-q\theta_1).
\end{align}    
We call $ g $ the binding density. 

If we choose the circular uniform density, which is the most diffuse circular distribution, as the binding density, then the circula densty becomes the independent circula, that is, $ c(\theta_1, \theta_2)=(2 \pi)^{-2} $. Conversely, if we choose a highly concentrated binding density, the circula density produces highly dependent structure between two circular random variables. This means the resultant length of the binding density $ g $ can be used as a mesure of dependence of the circula $ c $ defined in (\ref{eq:construction a pair circula}). Refer to Section 2.3 of \citet{jones2015class} for the relationships between the resultant length of the binding density and other pre-exisitng dependence measures of circular random variables.

%\section{Simulation study}
%We see how the approach works in estimating the dependence structure thorough a simulation study. 
%
%We generate cross-sectionary and serialy independent process with dimension $ m=2 $ and length $ T=300 $, from the Wrapped Cauchy distribution with mean direction $ \mu=\pi/2 $ and concent parameter $ \rho=0.5 $. 

\section{Real data analysis}\label{sec:Real data analysis}
\begin{figure}
%    \centering
    \includegraphics[height=4cm]{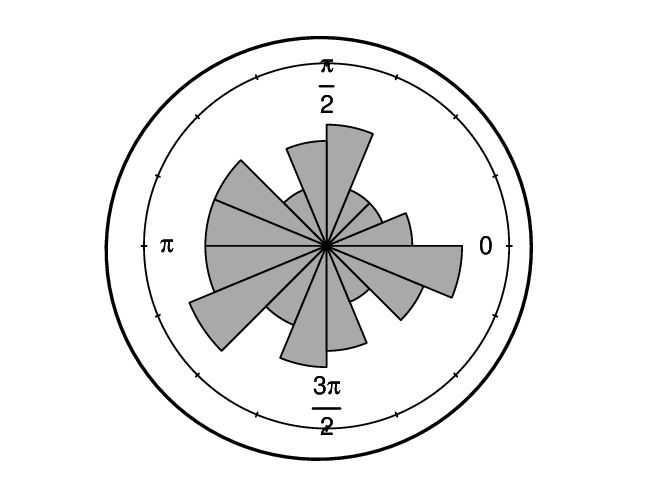}
    \includegraphics[height=4cm]{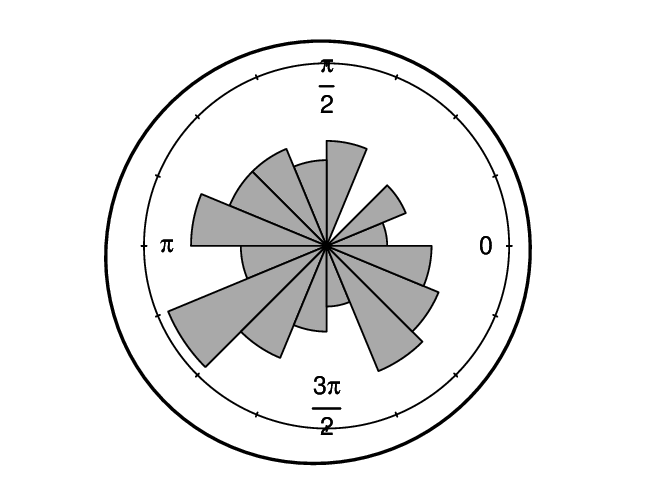}
    \includegraphics[height=4cm]{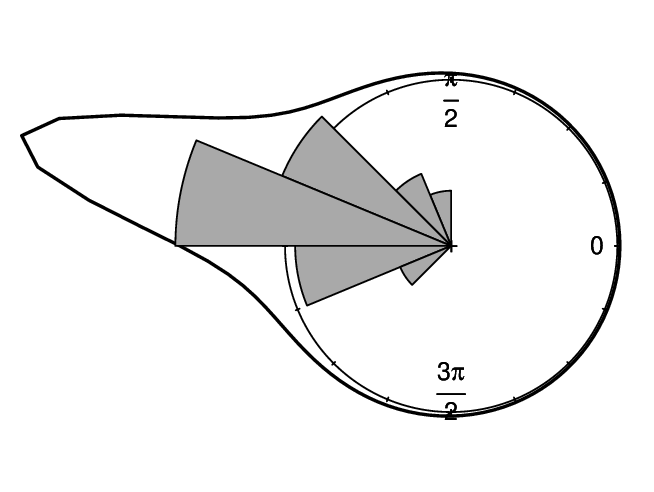}% \\
    \caption{Histograms and fitted circular density functions in polar representation. (Left) 1. Pine Grove. (Center) 2. Hood River. (Right) 3. Brookings. All figures are for hourly data.}
    \label{fig:rose diagram and fitted density}
\end{figure}

We fit the multivariate circular stationary Markov process with order $ p=2 $ to wind direction data recorded in Pine Grove, Hood River and Brookings in USA. The data are available from the United States Bureau of Reclamation website\footnote{https://www.usbr.gov/pn/agrimet/webagdayread.html.}. We obtained hourly and quarter-hourly data with the period 6th Feb. 2015 - 7th Feb. 2015 (two days). The numbers of time points are $ T=48 $ for hourly data and $ T=192 $ for quarter-hourly data, respectively. Both are three dimensional data $ (m=3) $, therefore, total numbers of observations are $ N=144 $ for hourly data and $ N=576 $ for quarter-hourly data, respectively. The excerpts of these data are shown in Tables \ref{tab: Hourly wind direction data}-\ref{tab: 15 minute wind direction data}. The  histograms (rose diagrams) of the data are given in Figure \ref{fig:rose diagram and fitted density}. The locations of the above three weather stations are given in Table \ref{tab: locations}. Note that Pine Grove and Hood River are very close each other.

\begin{table}[h]
	\centering
	\begin{tabular}{cc|ccc}
%		\hline
 Time index & Time & 1. Pine Grove & 2. Hood River & 3. Brookings \\ 
		\hline
  1 & Feb. 6 00:00 & 6.2169 & 2.4155 & 2.6791 \\ 
  2 & Feb. 6 01:00 & 5.2639 & 3.9305 & 2.8257 \\ 
  3 & Feb. 6 02:00 & 1.5429 & 0.4655 & 2.8327 \\ 
  4 & Feb. 6 03:00 & 6.1331 & 5.8119 & 2.8938 \\ 
  5 & Feb. 6 04:00 & 3.5448 & 3.8991 & 2.8571 \\ 
  6 & Feb. 6 05:00 & 2.1450 & 2.3562 & 3.1870 \\ 
		$ \vdots $&$ \vdots $&$ \vdots $&$ \vdots $&$ \vdots $ \\
  43 & Feb. 7 18:00 & 1.3898 & 1.2634 & 3.4854 \\ 
  44 & Feb. 7 19:00 & 1.3409 & 2.1886 & 2.8135 \\ 
  45 & Feb. 7 20:00 & 0.5227 & 1.2137 & 2.9252 \\ 
  46 & Feb. 7 21:00 & 4.7665 & 5.9324 & 1.9827 \\ 
  47 & Feb. 7 22:00 & 4.4070 & 2.5709 & 1.9076 \\ 
  48 & Feb. 7 23:00 & 3.8153 & 4.2220 & 2.0769 \\ 
%		\hline
	\end{tabular}
\caption{Hourly wind direction data (radians)}\label{tab: Hourly wind direction data}
\end{table}

\begin{table}[h]
	\centering
	\begin{tabular}{cc|ccc}
%		\hline
Time index & Time & 1. Pine Grove & 2. Hood River & 3. Brookings \\ 
		\hline
1 & Feb. 6  00:00 & 6.2169 & 2.4155 & 2.6791 \\ 
2 & Feb. 6  00:15 & 6.1261 & 0.0641 & 2.7855 \\ 
3 & Feb. 6  00:30 & 4.9253 & 0.0134 & 2.7943 \\ 
4 & Feb. 6  00:45 & 3.1032 & 3.4854 & 2.8641 \\ 
5 & Feb. 6  01:00 & 5.2639 & 3.9305 & 2.8257 \\ 
6 & Feb. 6  01:15 & 4.2010 & 3.4278 & 2.8047 \\ 
$ \vdots $&$ \vdots $&$ \vdots $&$ \vdots $&$ \vdots $ \\
187 & Feb. 7 22:30 & 0.5505 & 2.6861 & 1.7541 \\ 
188 & Feb. 7 22:45 & 4.0457 & 2.6686 & 1.8884 \\ 
189 & Feb. 7 23:00 & 3.8153 & 4.2220 & 2.0769 \\ 
190 & Feb. 7 23:15 & 2.6930 & 0.9488 & 2.0961 \\ 
191 & Feb. 7 23:30 & 5.0894 & 2.1398 & 2.1328 \\ 
192 & Feb. 7 23:45 & 0.3925 & 2.9060 & 2.2951 \\ 
%		\hline
	\end{tabular}
\caption{Quarter-hourly wind direction data (radians)}\label{tab: 15 minute wind direction data}
\end{table}

\begin{table}
	\centering
\begin{tabular}{|c|c|c|c|}
	\hline 
	& 1. Pine Grove & 2. Hood River & 3. Brookings \\ 
	\hline 
	Latitude &  $ 45.65222 $ N & $ 45.68444 $ N & $ 42.03 $ N \\ 
	\hline 
	Longitude & $ -121.50916 $ W & $ -121.51805 $ W & $ -124.24083 $ W \\ 
	\hline 
\end{tabular}  
\caption{Locations of the three weather stations}\label{tab: locations}
\end{table}

We fit the model (\ref{eq:circular joint density}) with the circula density $ c(\bm{\upsilon}) $ in (\ref{eq:pair circulas decomposition in stationary Markov}). Concretely, it can be written down
\begin{align*}
	f(\bm{\phi})=(2\pi)^N
	&\left\{
	 \prod_{t=1}^{T} \prod_{i=a(t)+1}^{b(t)} \prod_{j=a(t)}^{i-1}
	 (2\pi)^2c_{\ell_1,\ell_{2,0},0}^\ast \bigl(2\pi F(\upsilon_i|\bm{\upsilon}_{i-1:j+1}), 2\pi F(\upsilon_j|\bm{\upsilon}_{i-1:j+1})\bigl) 
	\right\} \\
	&	\left\{
	\prod_{t=2}^{T} \prod_{k=1}^{\min(t-1,p)} \prod_{i=a(t)}^{b(t)} \prod_{j=a(t-k)}^{b(t-k)}
	(2\pi)^2c_{\ell_1,\ell_{2,k},k}^\ast \bigl(2\pi F(\upsilon_i|\bm{\upsilon}_{i-1:j+1}), 2\pi F(\upsilon_j|\bm{\upsilon}_{i-1:j+1})\bigl) 
	\right\} \\
	& \prod_{j=1}^m \prod_{t=1}^T f_j(\theta_{j,t})
\end{align*}
where $ \ell_1=i-m(t-1) $ and $ \ell_{2,k}=j-m(t-k-1) $. Here we set $p=2$. After all, the functions to be specified are given in Table \ref{tab: functions to be specified}.
\begin{table}[h]
	\centering
	\begin{tabular}{ccc|c}
		%		\hline
		$ f_1 $  & $ f_2 $  & $ f_3 $  & marginals  \\ 
		\hline
		$ c_{2,1,0}^\ast $ & $ c_{3,1,0}^\ast $   & $ c_{3,2,0}^\ast $& controling cross sectional dependence  \\ 
		\hline
		$ c_{1,1,1}^\ast $ & $ c_{1,2,1}^\ast $   & $ c_{1,3,1}^\ast $&   \\ 
		$ c_{2,1,1}^\ast $ & $ c_{2,2,1}^\ast $   & $ c_{2,3,1}^\ast $& controling serial  dependence (lag1) \\ 
		$ c_{3,1,1}^\ast $ & $ c_{3,2,1}^\ast $   & $ c_{3,3,1}^\ast $&  \\ 
		\hline
		$ c_{1,1,2}^\ast $ & $ c_{1,2,2}^\ast $   & $ c_{1,3,2}^\ast $&   \\ 
		$ c_{2,1,2}^\ast $ & $ c_{2,2,2}^\ast $   & $ c_{2,3,2}^\ast $& controling serial  dependence (lag2) \\ 
		$ c_{3,1,2}^\ast $ & $ c_{3,2,2}^\ast $   & $ c_{3,3,2}^\ast $&  \\ 
	\end{tabular}
	\caption{Functions to be specified}\label{tab: functions to be specified}
\end{table}
The construction of pair circula densities is done by (\ref{eq:construction a pair circula}) with $ q=1 $. Here we use the wrapped Cauchy distribution whose location parameter is zero for binding densities. That is,
\begin{align*}
	c_{\ell_1,\ell_{2,k},k}^\ast \bigl(\theta_1, \theta_2 \bigl)
	=\frac{1}{2 \pi} \, \textrm{wC}(\theta_2-\theta_1;0, \rho_{\ell_1,\ell_{2,k},k})
\end{align*}
where
\begin{align*}
\textrm{wC}(\theta;\mu, \rho)
=\frac{1}{2 \pi} \frac{1-\rho^2}{1+\rho^2-2\rho \cos(\theta-\mu)}.
\end{align*}
For marginal distributions, we fit the wrapped Cauchy distribution with possibly non-zero location parameter:
\begin{align*}
f_j(\theta)=\textrm{wC}(\theta;\mu_j,\rho_j).
\end{align*}
The reason of this choice lies in its representability of the circular distribution function. The main burden in computing the pair circulas is the evaluation of the arguments, which include the circular distribution function. Unfortunately, many of preexisting circular models have no analytical forms of the distribution functions, and the wrapped Cauchy is one of the few exceptions. Due to \citet{F95}, the distribution function of the wrapped Cauchy is given by
\begin{align*}
\frac{1}{2\pi} \arccos
\left( 
\frac{(1+\rho^2)\cos(\theta-\mu)-2\rho}{1+\rho^2-2\rho\cos(\theta-\mu)}
\right)
\end{align*}

We estimate all the parameters by Markov chain Monte Carlo (MCMC) method. The setting for MCMC estimation is given in Table \ref{tab: MCMC setting} and the summary of the posterior distributions for hourly data and quarter-hourly data are given in Table \ref{tab: posterior}.

\begin{table}
	\centering
\begin{tabular}{|c|c|c|c|c|}
	\hline 
	 Chains & Iteration & Warmup & Thinning & Prior \\ 
	\hline 
	 3 & 3,000 & 100 & 1 & Non informative \\ 
	\hline  
\end{tabular}  
\caption{The setting for MCMC estimation}\label{tab: MCMC setting}
\end{table}

\begin{table}[h]
\centering
\begin{tabular}{rrrrr|rrr}			
\hline
&& \multicolumn{3}{c|}{Hourly} & \multicolumn{3}{c}{Quarter-hourly} \\
&& \multicolumn{1}{c}{mean} & \multicolumn{1}{c}{sd} & \multicolumn{1}{c|}{median} & \multicolumn{1}{c}{mean} & \multicolumn{1}{c}{sd} & \multicolumn{1}{c}{median} \\ 
\hline
&$ \mu_{1} $ & 3.4687 & 1.0307 & 3.5141 & 3.7353 & 1.0635 & 3.8258 \\ 
&$ \mu_{2} $ & 3.7930 & 0.7829 & 3.8284 & 3.5661 & 0.7520 & 3.5586 \\ 
margi-&$ \mu_{3} $ & 2.8958 & 0.0407 & 2.8952 & 2.9306 & 0.0213 & 2.9306 \\ \cdashline{2-8}
nals&$ \rho_{1} $ & 0.1395 & 0.0911 & 0.1297 & 0.0710 & 0.0461 & 0.0656 \\ 
&$ \rho_{2} $ & 0.1859 & 0.0964 & 0.1864 & 0.0965 & 0.0517 & 0.0954 \\ 
&$ \rho_{3} $ & 0.8260 & 0.0308 & 0.8280 & 0.8270 & 0.0152 & 0.8275 \\ 
\hline
&$ \rho_{12,0} $ & 0.5488 & 0.0718 & 0.5581 & 0.3692 & 0.0421 & 0.3700 \\ 
lag 0&$ \rho_{13,0} $ & 0.0096 & 0.0074 & 0.0081 & 0.0062 & 0.0037 & 0.0058 \\ 
&$ \rho_{23,0} $ & 0.0371 & 0.0227 & 0.0367 & 0.0061 & 0.0037 & 0.0057 \\ 
\cdashline{2-8}
&$ \rho_{11,1} $ & 0.0886 & 0.0696 & 0.0731 & 0.1853 & 0.0490 & 0.1854 \\ 
&$ \rho_{12,1} $ & 0.2610 & 0.1141 & 0.2611 & 0.2736 & 0.0476 & 0.2746 \\ 
&$ \rho_{13,1} $ & 0.0108 & 0.0094 & 0.0082 & 0.0039 & 0.0031 & 0.0031\\ 
&$ \rho_{21,1} $ & 0.2271 & 0.0876 & 0.2283 & 0.1546 & 0.0557 & 0.1557 \\ 
lag 1&$ \rho_{22,1} $& 0.0871 & 0.0615 & 0.0783 & 0.3103 & 0.0490 & 0.3103 \\ 
&$ \rho_{23,1} $ & 0.0085 & 0.0078 & 0.0063 & 0.0024 & 0.0020 & 0.0019 \\ 
&$ \rho_{31,1} $ & 0.0779 & 0.0649 & 0.0619 & 0.0622 & 0.0420 & 0.0561 \\ 
&$ \rho_{32,1} $ & 0.0760 & 0.0632 & 0.0593 & 0.0450 & 0.0334 & 0.0383 \\ 
&$ \rho_{33,1} $ & 0.8637 & 0.0318 & 0.8671 & 0.9283 & 0.0073 & 0.9288 \\ 
\cdashline{2-8}
&$ \rho_{11,2} $ & 0.1055 & 0.0756 & 0.0916 & 0.0234 & 0.0206 & 0.0178 \\ 
&$ \rho_{12,2} $ & 0.0896 & 0.0662 & 0.0766 & 0.0722 & 0.0437 & 0.0673 \\ 
&$ \rho_{13,2} $ & 0.0555 & 0.0486 & 0.0419 & 0.0375 & 0.0291 & 0.0310 \\ 
&$ \rho_{21,2} $ & 0.1730 & 0.1057 & 0.1633 & 0.0482 & 0.0348 & 0.0419 \\ 
lag 2&$ \rho_{22,2} $ & 0.1010 & 0.0755 & 0.0866 & 0.0447 & 0.0334 & 0.0384 \\ 
&$ \rho_{23,2} $ & 0.2557 & 0.1137 & 0.2518 & 0.0330 & 0.0263 & 0.0270 \\ 
&$ \rho_{31,2} $ & 0.2181 & 0.1079 & 0.2129 & 0.0544 & 0.0377 & 0.0484 \\ 
&$ \rho_{32,2} $ & 0.1026 & 0.0752 & 0.0876 & 0.0466 & 0.0340 & 0.0403 \\ 
&$ \rho_{33,2} $ & 0.1021 & 0.0787 & 0.0859 & 0.0595 & 0.0399 & 0.0538 \\ 
\hline
\end{tabular}
\caption{Summary of posterior distributions}\label{tab: posterior}
\end{table}

Let us use a posterior mean as an estimator of the parameters. For lag 0, the posterior mean of $\rho_{12,0}$ is much larger than those of $\rho_{13,0}$ and $\rho_{23,0}$ for both hourly and quarter-hourly data. This implies the cross-sectional dependence between 1. Pine Grove and 2. Hood River are much higher than those between 1. Pine Grove and 3. Brookings, and between 2. Hood River and 3. Brookings. Considering the closeness between 1. Pine Grove and 2. Hood River, the result is convincing. For lag 1, the significant feature is the strong auto-dependence of 3. Brookings, implied by large posterior means of $\rho_{33,1}$ for both hourly and quarter-hourly data. For lag 2, although the posterior mean of $\rho_{23,2}$ for hourly data is somewhat large, we cannot find strong auto-dependence for both hourly and quarter-hourly data.

For marginal distributions, the posterior means of $\rho_3$ are larger than those of $\rho_1$ and $\rho_2$ for both hourly and quarter-hourly data. This implies the data of 3. Brookings is more concentrated to its mean direction than those of 1. Pine Grove and 2. Hood River. For hourly data, the fitted wrapped Cauchy density functions and histograms in polar representation are displayed in Figure \ref{fig:rose diagram and fitted density}. All seem to fit to the data well. 

\section*{Appendix}
The $ N $-dimensional circular density is decomposed as
\begin{align}\label{eq:joint circula}
c(\bm{\upsilon})
=c(\upsilon_1) \prod_{i=2}^{N}  \frac{c(\bm{\upsilon}_{i:1})}{c(\bm{\upsilon}_{i-1:1})}.
\end{align}
For $ i > j $, we have
\begin{align*}
\frac{c(\bm{\upsilon}_{i:j})}{c(\bm{\upsilon}_{i-1:j})} 
=\frac{c(\bm{\upsilon}_{i:j})}{c(\bm{\upsilon}_{i-1:j+1})}
\frac{c(\bm{\upsilon}_{i-1:j+1})}{c(\bm{\upsilon}_{i-1:j})}
=f(\upsilon_{i},\upsilon_{j} | \bm{\upsilon}_{i-1:j+1})
\frac{1}{f(\upsilon_j | \bm{\upsilon}_{i-1:j+1})}
\end{align*}
The conditional bivariate circula density $ f(\upsilon_{i},\upsilon_{j} | \bm{\upsilon}_{i-1:j+1}) $ is reexpressed with another pair circula density $ c^\ast_{i,j} $ as
\begin{align*}
f(\upsilon_{i},\upsilon_{j} | \bm{\upsilon}_{i-1:j+1})
=(2\pi)^2 c^\ast_{i,j}\bigl(2\pi F(\upsilon_{i} | \bm{\upsilon}_{i-1:j+1}), 2\pi F(\upsilon_{j} | \bm{\upsilon}_{i-1:j+1})\bigl) f(\upsilon_{i} | \bm{\upsilon}_{i-1:j+1}) f(\upsilon_{j} | \bm{\upsilon}_{i-1:j+1}). 
\end{align*}
Therefore, we have the expression  
\begin{align*}
\frac{c(\bm{\upsilon}_{i:j})}{c(\bm{\upsilon}_{i-1:j})} 
=&(2\pi)^2 c^\ast_{i,j}\bigl(2\pi F(\upsilon_{i} | \bm{\upsilon}_{i-1:j+1}), 2\pi F(\upsilon_{j} | \bm{\upsilon}_{i-1:j+1})\bigl) f(\upsilon_{i} | \bm{\upsilon}_{i-1:j+1}) \\
=&(2\pi)^2 c^\ast_{i,j}\bigl(2\pi F(\upsilon_{i} | \bm{\upsilon}_{i-1:j+1}), 2\pi F(\upsilon_{j} | \bm{\upsilon}_{i-1:j+1})\bigl) \frac{c(\bm{\upsilon}_{i:j+1})}{c(\bm{\upsilon}_{i-1:j+1})},
\end{align*}
Recursive calculation, and $ c(\bm{\upsilon}_{i-1:i-1})=c(\upsilon_{i})=(2\pi)^{-1} $ leads to
\begin{align*}
\frac{c(\bm{\upsilon}_{i:1})}{c(\bm{\upsilon}_{i-1:1})} 
=(2\pi)^{2i-3} \prod_{j=1}^{i-1} c^\ast_{i,j}\bigl(2\pi F(\upsilon_{i} | \bm{\upsilon}_{i-1:j+1}), 2\pi F(\upsilon_{j} | \bm{\upsilon}_{i-1:j+1})\bigl).
\end{align*}
Substituting this into (\ref{eq:joint circula}) together with $ c(\upsilon_1)=(2\pi)^{-1} $, we have the expression (\ref{eq:pair circulas decomposition}). 

\bibliography{ogata.bib}   % name your BibTeX data base
\bibliographystyle{chicago}

\end{document}